\documentclass{emulateapj}

\usepackage{amsmath}
\usepackage{amssymb}
\usepackage{graphicx}
\usepackage{color}
\usepackage[colorlinks=false]{hyperref}
\usepackage{color}
\hypersetup{
    colorlinks=false,
    pdfborder={0 0 0},
}
\usepackage{epstopdf}

\def\gtaprx {\lower .1ex\hbox{\rlap{\raise .6ex\hbox{\hskip .3ex
    {\ifmmode{\scriptscriptstyle >}\else
        {$\scriptscriptstyle >$}\fi}}}
    \kern -.4ex{\ifmmode{\scriptscriptstyle \sim}\else
        {$\scriptscriptstyle\sim$}\fi}}}
\def\ltaprx {\lower .1ex\hbox{\rlap{\raise .6ex\hbox{\hskip .3ex
    {\ifmmode{\scriptscriptstyle <}\else
        {$\scriptscriptstyle <$}\fi}}}
    \kern -.4ex{\ifmmode{\scriptscriptstyle \sim}\else
        {$\scriptscriptstyle\sim$}\fi}}}
\newcommand{\note}[1]{\emph{\textcolor{red}{}}}

\newcommand{\Ms}{{\ensuremath{{M}_{\odot} }}}

\newcommand{\Ni}{{\ensuremath{^{56}\mathrm{Ni}}}}

\newcommand{\jFe}{{\ensuremath{^{54}\mathrm{Fe}}}}
\newcommand{\iFe}{{\ensuremath{^{52}\mathrm{Fe}}}}
\newcommand{\Co}{{\ensuremath{^{56}\mathrm{Co}}}}

\newcommand{\Hea}{{\ensuremath{^{3} \mathrm{He}}}}
\newcommand{\Heb}{{\ensuremath{^{4} \mathrm{He}}}}
\newcommand{\Hy}{{\ensuremath{^{1} \mathrm{H}}} }
\newcommand{\Ox}{{\ensuremath{^{16}\mathrm{O}}}}
\newcommand{\Ti}{{\ensuremath{^{44}\mathrm{Ti}}}}
\newcommand{\Si}{{\ensuremath{^{28}\mathrm{Si}}}}
\newcommand{\Mg}{{\ensuremath{^{24}\mathrm{Mg}}}}

\newcommand{\Cx}{{\ensuremath{^{12}\mathrm{C}}}}

\newcommand{\Cr}{{\ensuremath{^{48}\mathrm{Cr}}}}
\newcommand{\Ca}{{\ensuremath{^{40}\mathrm{Ca}}}}

\newcommand{\Ar}{{\ensuremath{^{36}\mathrm{Ar}}}}
\newcommand{\Sx}{{\ensuremath{^{32}\mathrm{S}}}}
\newcommand{\Nx}{{\ensuremath{^{14}\mathrm{N}}}}
\newcommand{\Ne}{{\ensuremath{^{20}\mathrm{Ne}}}}

\newcommand{\na}{{New Astronomy}}%

\newcommand{\FIGFF}[2]{{\ref{fig:#2}{#1}}}

\newcommand{\FIG}[2]{{Fig.~\FIGFF{#1}{#2}}}
\newcommand{\Fig}[1]{{\FIG{}{#1}}}

\newcommand{\cm}{{\ensuremath{\mathrm{cm}}}}
\newcommand{\erg}{{\ensuremath{\mathrm{erg}}}}

\newcommand{\CASTRO}{\texttt{CASTRO}}
\newcommand{\KEPLER}{\texttt{KEPLER}}
\newcommand{\STELLA}{\texttt{STELLA}}

\newcommand{\gcc}{\ensuremath{\mathrm{g}\,\mathrm{cm}^{-3}}}

\newcommand{\ken}[1]{\textcolor{black}{#1}}

\makeatletter

\newcommand{\Rmnum}[1]{\expandafter\@slowromancap\romannumeral #1@}
\makeatother

\begin{document}

\title{Magnetar-Powered Supernovae in Two Dimensions. II. Broad-Line Supernovae 
I\small{c}} 

\author{Ke-Jung Chen\altaffilmark{1,2,3*}, 
            Takashi J. Moriya\altaffilmark{1}, 
            Stan Woosley\altaffilmark{3}, 
            Tuguldur Sukhbold\altaffilmark{3}, 
            Daniel J. Whalen\altaffilmark{4}, 
            Yudai Suwa\altaffilmark{5}, and 
            Volker Bromm\altaffilmark{6}
            } 

\altaffiltext{1}{Division of Theoretical Astronomy, National Astronomical Observatory of 
Japan, Tokyo 181-8588, Japan} 
\altaffiltext{2}{Institute of Astronomy and Astrophysics, Academia Sinica,  Taipei 10617, 
Taiwan}
\altaffiltext{3}{Department of Astronomy \& Astrophysics, University of California, Santa 
Cruz, CA 95064, USA}
\altaffiltext{4}{Institute of Cosmology and Gravitation, Portsmouth University, Portsmouth, 
UK}
\altaffiltext{5}{Center for Gravitational Physics, Yukawa Institute for Theoretical Physics, Kyoto University, Oiwake-cho, Kitashirakawa, Sakyo-ku, Kyoto, 606-8502, Japan} 
\altaffiltext{6}{Department of Astronomy, University of Texas, Austin, TX 78712, USA} 
\altaffiltext{*}{EACOA Fellow, email: {\tt ken.chen@nao.ac.jp}}

\begin{abstract}

Nascent neutron stars with millisecond periods and magnetic fields in excess of $10^{16}$ 
Gauss can drive highly energetic and asymmetric explosions known as magnetar-powered 
supernovae.  These exotic explosions are one theoretical interpretation for supernovae  
\ken{Ic-BL} which are sometimes associated with long gamma-ray bursts.  Twisted 
magnetic field lines extract the rotational energy of the neutron star and release it as a disk 
wind or a jet with energies greater than 10$^{52}$ erg over $\sim 20$ sec.  What fractions 
of the energy of the central engine go into the wind and the jet remain unclear.  We have 
performed two-dimensional hydrodynamical simulations of magnetar-powered supernovae 
(SNe) driven by disk winds and jets with the \CASTRO\ code to investigate the effect of the 
central engine on nucleosynthetic yields, mixing, and light curves.  We find that these 
explosions synthesize less than 0.05 \Ms\ of \Ni\ and that this mass is not very sensitive to 
central engine type. The morphology of the explosion can provide a powerful diagnostic of 
the properties of the central engine. In the absence of a circumstellar medium these events 
are not very luminous, with peak bolometric magnitudes $M_b \sim -16.5 $ due to low \Ni\ 
production. 

\end{abstract}

  
\section{Introduction}

Most stars from 30 - 80 \Ms\ eventually collapse to black holes because the energy released 
by core collapse cannot drive a shock that is powerful enough to overcome the ram pressure 
of infall, so core bounce fails to produce an explosion \citep[e.g.][]{Suk15, Ert16}.  But this 
picture can change with rapidly rotating stars, in which a neutron star (NS) with a period of a 
few milliseconds may be born.  Rotation can amplify the magnetic field of the NS above  $10
^{15}$ G, creating a magnetar.  The magnetar might spin down quickly by magnetic braking 
and release its rotational energy in the form of a radiatively-dominated disk wind \citep{Dun92,
Tho93}.  During a brief, early phase of braking, the radiation can be in the form of x-rays and 
soft gamma-rays \citep{Kou98,Gae05,Woo06, Mae07,Mer08,Esp09}.   In some cases, if 
magnetorotational instabilities arise they can launch a collimated jet that pierces the outer 
layers of the star and produces a gamma-ray burst (GRB) \citep[e..g.][]{Leb70,Bur07,Uzd06,
Mos14}.  

If both a disk wind and jet are present, a highly asymmetric SN explosion may accompany 
the burst \ken{\citep[e..g.][]{Met11,Sok16}}.  Such events release energies of up to $\sim 10^{52} \erg$, 
10 times those of conventional core-collapse (CC) SNe \citep{Pac98, Iwa99}.  These 
magnetar-powered SNe are likely observed as broad-line Type Ic SNe (SNe Ic-BL), which 
are often referred to as hypernovae, and they are among the most energetic explosions in 
the Universe \citep{Smi13a}. \ken{SNe Ic-BL have a very broad absorption lines of oxygen and iron 
but lacking of helium and hydrogen in their spectra. Their light curves (LCs)  peak at absolute magnitude $\approx$ -18 $-$ -20 mag and the shape of LC is different from that of SNe Ic, which shows a broad peak and a slow tail \citep{Iwa00}. The ejecta of SNe Ic-BL expand at velocities about $30,000$ km/s which is much faster than that of normal SNe \citep{Vin15}.  About $0.05 - 0.3 \Ms$ \Ni\  is estimated to form in SNe Ic-BL \citep{Can13,Pre16}}.  Since their explosion engines are closely tied to their central 
remnants, SNe Ic-BL are promising candidates for studying the physics of compact objects 
because they may account for the SNe associated with some GRBs \citep{1998bw,
1998bw2,2010bh}.  Less extreme ($10^{14} - 10^{15}$ G,  $5-10$ ms) magnetars may 
explain superluminous SNe \citep[SLSNe][]{Woo10, Kas10, Che16}, \ken{ because a 
	substantial fraction of the total rotational energy of the neutron star is 
	emitted as light at late time. \cite{Che16} used 2D simulations to study the mangetar-powered SNe by neutron stars of 
a constant magnetic field strength of $4\times10^{14}$ G, with initial rotational periods of 1 ms 
and 5 ms. They found that fluid instabilities cause a strong mixing and fracture shells of ejecta 
into filamentary structures. The observational signatures of the resulting supernova could be very 
different from those predicted by the 1D models.}

Predicting the observational signatures of magnetar-powered SNe is a key to properly 
identifying them as more are discovered by the new SN factories such as the Palomar 
Transient Factory \citep[PTF;][]{ptf1}, the Panoramic Survey Telescope and Rapid 
Response System \citep[Pan-STARRS;][]{panstarrs} and the Large Synoptic Survey 
Telescope \citep[LSST;][]{lsst}.  Multidimensional simulations that bridge a large range of 
spatial scales are needed to model the light curves of magnetar-powered SNe because of
their inherent asymmetry.  Previous studies have mainly focused on just the physics of the 
central engines of these explosions \citep[e.g.,][]{Bur07} but larger-scale simulations of 
mixing in jet-powered SNe \citep{Cou11,Pap14a,Pap14b} and neutrino-wind driven SNe \citep{Jog10,Nor10,Won15} have now been done. In this paper we study magnetar-powered SNe Ic-BL 
driven by central engines that are combinations of jets and disk winds. Our two-dimensional
models include nucleosynthesis but not radiation transport, and they evolve the explosion
from early internal mixing to breakout and then homologous expansion in order to calculate 
their observational signatures.  In Section 2 we describe our progenitor models, explosion
simulations and light curve calculations.  The evolution of the explosions, including mixing, 
is examined in Section 3 and we discuss our results and conclude in Section 4.

\section{Numerical Setup}

We take as our initial conditions the collapsing carbon-oxygen core of a massive star that 
has been evolved from central carbon ignition to the onset of core collapse.  At this stage, 
profiles for the star are mapped into the \CASTRO{} code and exploded with a variety of 
central engines.  Blast profiles from \CASTRO{} are then evolved with the radiation 
hydrodynamics code \STELLA{} to obtain light curves for these events.

\subsection{\ken{\KEPLER} Progenitor Model}

The progenitor star in all our models is a non-rotating 10 \Ms\ carbon-oxygen core with 
initial \Cx\ and \Ox\ mass fractions of 0.172 and 0.828, respectively. This progenitor, taken 
from another simulation campaign \citep{Suk14}, roughly corresponds to the CO core of a 
non-rotating, solar-metallicity 35 \Ms\ star prior to the loss of its H and He envelope.  It is 
evolved from central carbon ignition until the beginning of collapse with the 1D stellar 
evolution code \KEPLER{} \citep{kepler}. The actual mass of the zero-age main sequence 
star can vary depending on mass loss, rotation, magnetic field, and other physical 
parameters since more than one combination of progenitor mass and physical parameters 
can lead to the same CO core mass.

We consider a stripped CO core because the SNe Ic-BL observed to date are likely the
explosions of massive stars that have shed their helium envelopes \citep{Iwa99,Nak01}.  
Such SNe must either be born with rapid rotation rates \citep{Yoo06} or be spun up at 
late times in a common envelope phase with another star (tidal locking) \citep{Fwh99, 
Izz04} or compact object \citep{zf01}.  Either case usually results in the expulsion of the 
hydrogen and helium envelopes in some type of outburst \citep[a luminous blue variable 
star, or LBV;][]{Bar01} or the ejection of a dense shell just before the death of the star. 
However, a large mass ejected from a massive star may reduce its angular momentum 
and prevent it from forming a rapidly rotating core, which is required for magnetar 
formation. An alternative path is quasi-chemically homogeneous evolution, which results 
in a large CO core that retains its angular momentum because there is no massive 
ejection.  Tidal locking in the case of close binaries could then provide additional angular 
momentum to create a rapidly rotating core.

\begin{figure}[h]
	\begin{center}
		\includegraphics[width=\columnwidth]{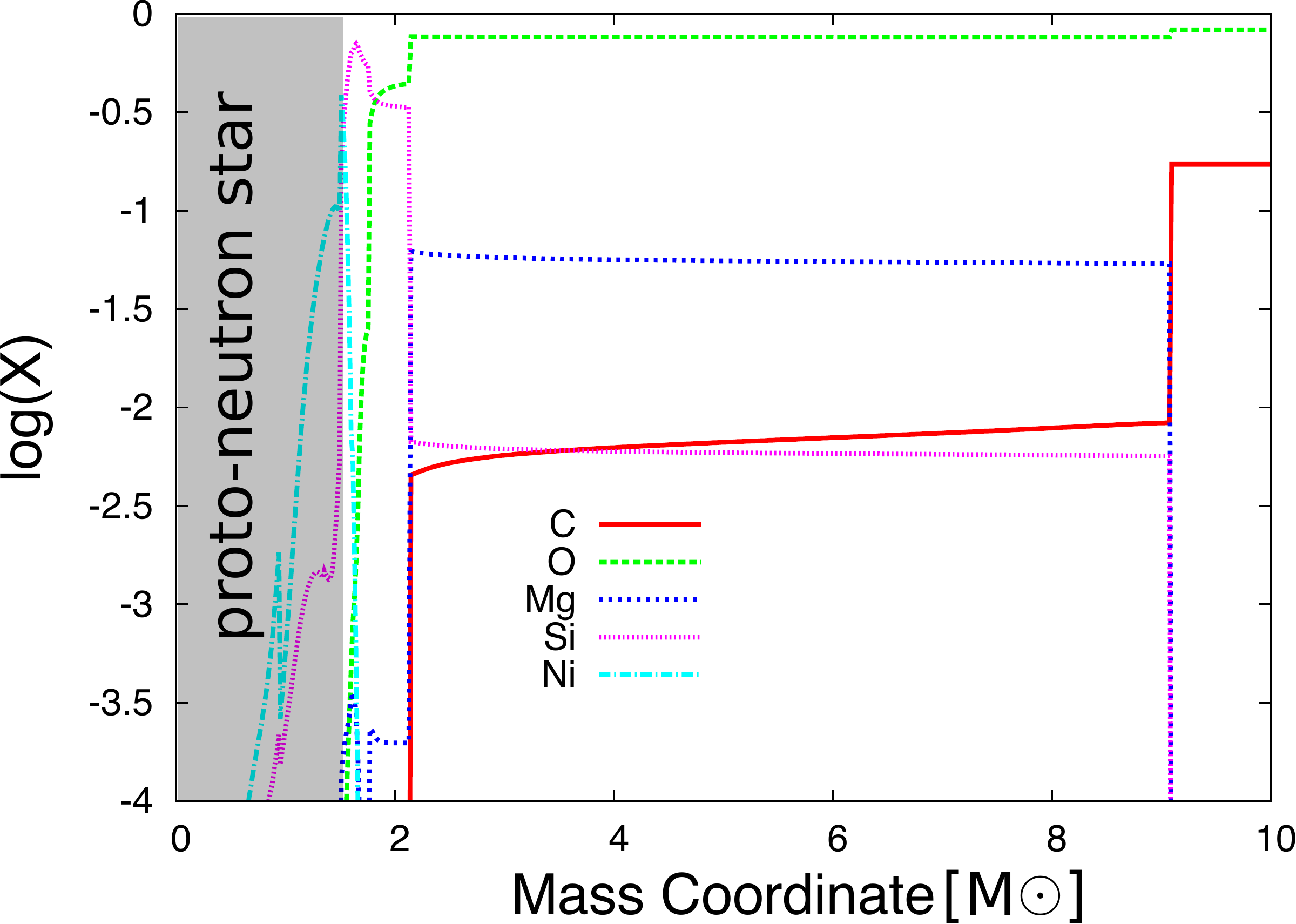} 
		\caption{Species mass fractions (top) for the 10 \Ms\ CO core 
			at the time of explosion \citep{Suk14}.  \ken{The edge of iron core defines as its infalling velocities reaching close to $10^8$ cm/s during its pre-supernova phase.} At this time,
			the iron core mass is 1.51 \Ms (shaded gray), and 
			it is assumed to collapse to a proto-neutron star which then becomes a magnetar.  \Ni\ at 
			mass coordinates below $\sim$ 1.6 \Ms\ would fall back directly onto the NS. Therefore, \Ni\ appears in the SN ejecta must be made during the explosion.			
			\label{fig:1D}}
	\end{center}
\end{figure}

Our model was evolved for 14,100 years to form a 1.51 \Ms\ iron core with a radius of 
1570 km. The 1D stellar evolution model was halted when any region of the star began 
to collapse faster than 1000 km s$^{-1}$.  At this point it was mapped into a 2D grid in 
\CASTRO.  The composition and velocity profiles of the pre-SN core are shown in 
\Fig{1D}. 


\begin{figure}[h]
	\begin{center}
		\includegraphics[width=\columnwidth]{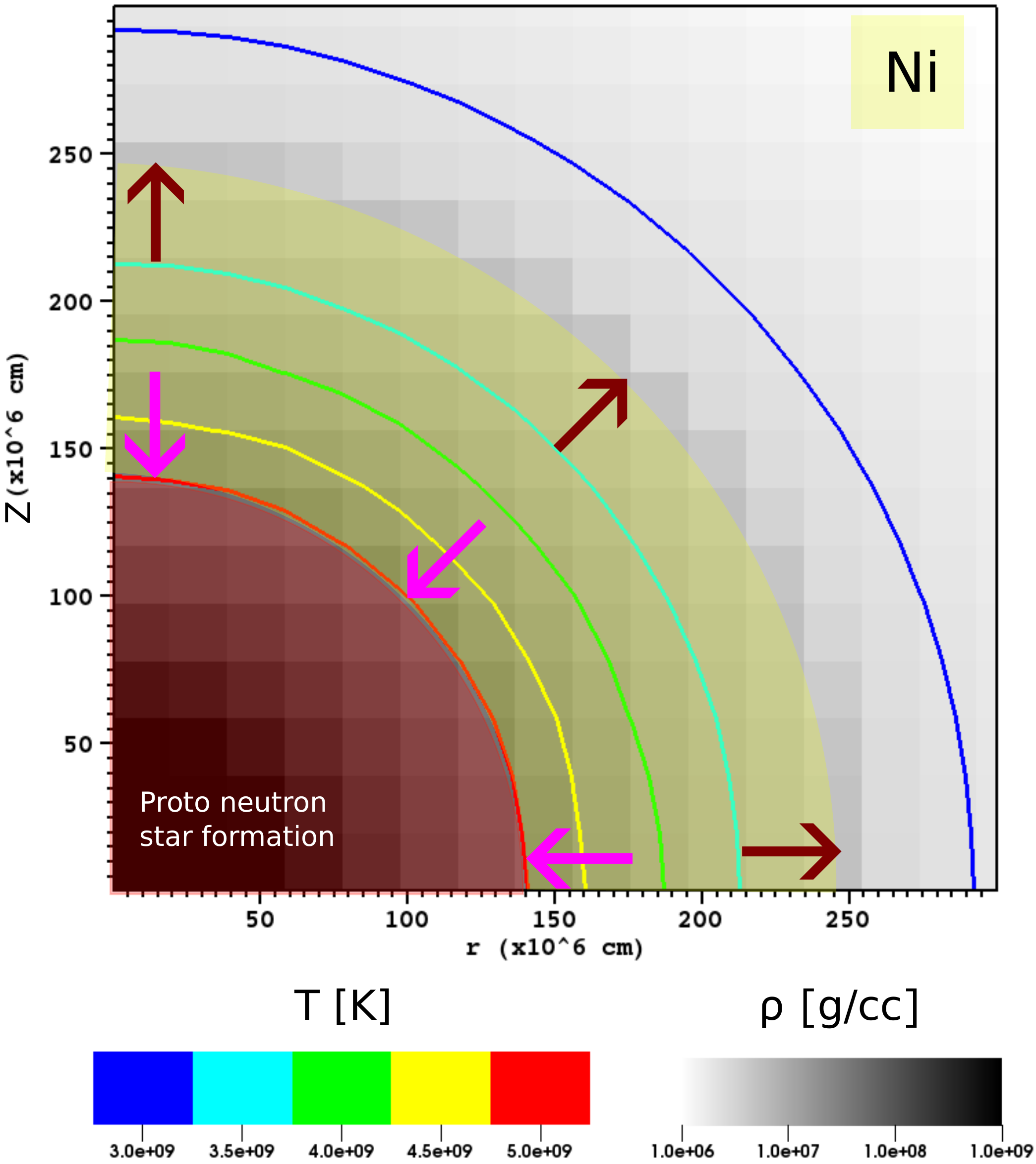} 
		\caption{Temperatures and densities in the 10 \Ms\ CO core prior to energy injection by the 
			magnetar.  Grey tones and color contours show the density and temperature, respectively.  
			The arrows roughly mark infalling gas and outgoing ejecta during the explosion. Yellow indicates 
			regions in which \Ni\ can be synthesized. Unless the magnetar forms immediately after the 
			explosion, most of the \Ni\ formed in the inner shell will fall back instead of being ejected.  
			\label{fig:p_ni}}
	\end{center}
\end{figure}

\subsection{2D \CASTRO\ Models}

\CASTRO\ is a multidimensional adaptive mesh refinement (AMR) radiation hydrodynamics 
code \citep{Alm10, Zha11} with an unsplit piecewise parabolic method hydro scheme \citep{
Woo84}.  \CASTRO\ has a Helmholtz equation of state based on \citet{Tim00}, which has 
relativistic electron-positron pairs of arbitrary degeneracy, ions (which are treated as an ideal 
gas) and photons.  We advect 17 isotopes, \Hy, \Hea, \Heb, \Cx, \Nx, \Ox, \Ne, \Mg, \Si, \Sx, 
\Ar, \Ca, \Ti, \Cr, \iFe, \jFe, \Ni, and use a simplified prescription for nuclear burning that takes 
all elements to be burned to \Ni\ when the gas temperature and density exceeds $4.3 \times
10^9$ K and $1\times10^6 \gcc$. Densities, velocities, temperatures and mass fractions from 
\KEPLER\ are mapped onto a 2D cylindrical grid in \CASTRO\ with the conservative scheme 
of \citet{Che13}, which preserves energies and masses over a large range of spatial scales.  
The mapping is done just before the formation of the NS, when infall velocities near the core 
reach $\sim 1000$ km s$^{-1}$.

We only simulate a quadrant of the star in 2D, so the mesh is $2 \times 10^{12}$ cm in both 
$r$ and $z$, or about twenty times the radius of the star, $r_* \sim 1.1 \times 10^{11}$ cm.  
The star is shrouded by a low-density envelope $\rho \propto r^{-3.1}$ that prevents mixing
as the forward shock plows up the circumstellar medium (CSM) after breakout.  The root 
grid has 256$^2$ zones and up to eight levels of refinement for an additional factor of up to 
256 (2$^8$) in spatial resolution.  The grid is refined on gradients in density, velocity, and 
pressure. This approach provides an effective simulation domain of $65,536 \times 65,536$ 
zones. 

The explosion energy from the magnetar is injected by hand.  We center eight nested grids
on the site of energy injection, each of which has twice the resolution of the grid above it for 
a maximum resolution equal to that of the lowest level of the AMR hierarchy. Reflecting and 
outflow boundary conditions are imposed on the inner and outer boundaries in both $r$ and 
$z$, respectively.  We use a monopole approximation for self-gravity, in which the 
gravitational potential is constructed from the radial average of the density and used to 
calculate gravity forces everywhere in the AMR hierarchy by linear interpolation.  This 
approximation is efficient and valid because the star is nearly spherically symmetric.  Point 
source gravity due to the compact remnant is also included in our models. 

We estimate the explosion energy and its timescale as follows.  The radius of the NS is 
assumed to be $R_n \sim10^{6}$ cm and the moment of inertia for a typical NS is $I \sim10
^{45}$ g cm$^2$ so its initial rotational energy is
\begin{equation}
E = \frac{1}{2}I\omega^2\approx 2\times 10^{52}P_{\rm ms}^{-2} \quad
\erg,
\label{eq:1}
\end{equation}
where $P_{ms}$ is the period of the magnetar in milliseconds. This energy can be released 
through dipole radiation,
\begin{equation}
\begin{split}
\frac{dE}{dt} & = -\frac{32\pi^4}{3c^3} (BR_n^3\sin\alpha)^2P^{-4}
\\ & \approx 10^{49}B_{15}^2 P_{\rm ms}^{-4} \quad \mathrm{erg~s^{-1}},
\end{split}
\label{eq:2}
\end{equation}
where $B_{15}= B/10^{15}$G and assume $\sin\alpha = 1$ for simplicity.  The spin-down 
time scale can be approximated as 
\begin{equation}
\tau_{d} \approx E/|\frac{dE}{dt}| \\  \approx 2000 B_{15}^{-2} P_{\rm ms}^{2} \quad \sec.
\end{equation}
If a magnetar forms with a rotation period of 1 ms and magnetic field stress of $10^{16}$ G 
a total energy of $\sim 2 \times10^{52}$ erg can be deposited into the surrounding core in 
20 sec in the form of a very energetic disk wind or a collimated jet after the formation of the 
NS.  Since the specifics of the central engine are not known, we consider four cases by 
varying the fraction of the energy that goes into the wind (isotropic) or the jet (anisotropic).  
\ken{ Both observational evidence \citep{Gae06,You16} and theoretical studies \citep{Pro98,Pro00} of the disk wind from the accretion disk suggest that such wind can be highly inhomogeneous and anisotropic. Because our simulation cannot resolve  the central disk, yet do not include MHD. Therefore, the mixing from our spherical  disk wind model can be the lower limit of actual mixing. }

Engine A is a purely jet-driven explosion, 
with all the energy going into a jet with a half opening angle $\phi=2.5^{\circ}$.  
\ken{ This model is a practical example mentioned in  \citep{Gil16} before.}
Engine B is 
a wind driven explosion in which all the energy of the explosion is deposited isotropically 
into the surrounding core. Engine C is both a jet and a wind in which 90\% of the energy is 
isotropic and 10\% goes into a jet with $\phi =2.5^{\circ}$.  Engine D is the same as Engine 
C but with $\phi=10^{\circ}$. Engines A and B are the two extreme cases while Engine C is 
motivated by the conventional GRB SN and Engine D represents the case of a wobbling jet. 
The precession is caused by kink instabilities which may often occur in core-collapse jets
in \citep{Bro16}. Such instabilities disperse the energy in a larger opening angle, so the jet may die before reaching the stellar surface without producing a 
GRB. \ken{ \cite{Gil16b} also suggested that a wide jet can form  if there are non-axisymmetric patterns in the core-collapse.} We summarize the four engines in Table~\ref{tbl:model}. \ken{These  models are
associated with the jet-feedback mechanism (JFM) of CCSNe \citep{Pap11, Lop13}. In JFM scenario, the jets launched from the central compact object must be fast and narrow, they would deposit their energy inside the star  through shock waves, then forming two hot bubbles to push out infalling 
gas. Therefore, the jet becomes slow, massive, and wide.  If the jet feedback is effective, accretion would 
halt on early and result in regular supernova explosions. Otherwise, the accretion lasts longer and it supply more energy to the jet and eventually creates much energetic explosions \citep{Sok16}.}

We assume that the wind and jet deposit their energy uniformly in a region 2,000 - 5,000 km from 
the center of the star.  In the wind, all of the energy is deposited as internal energy of the 
gas.  In the jet, $50\%$ of the energy goes into internal energy and the rest goes into 
kinetic energy by injecting a highly relativistic momentum flux with a speed of $1.5 \times 
10^{10}$  cm s$^{-1}$ along the polar axis.  In some cases the jet can blow out so much 
gas that extremely low densities result, which can cause numerical difficulties in the runs. 
To prevent blowout from creating a complete vacuum in the energy injection region we 
added a total mass of $5 \times 10^{-8}$ \Ms\ to these zones, which is completely 
negligible in comparison to the mass of star. 

Gas falling into the central 2,000 km of the grid is assumed to accrete onto the NS or black 
hole (BH).  364 grids at the deepest level of refinement resolve the injection region during
the simulation.  In each case, the shock is evolved until it reaches the outer grid boundary 
at $\sim$ 20 $r_*$, when the ejecta are expanding homologously. Due to the large number 
of levels of refinement, the four models took about 800,000 CPU hours on $\it Hopper$ at 
the National Energy Research Scientific Computing Center (NERSC).


\begin{deluxetable*}{cccccc}
	\tablecaption{Magnetar-powered Explosions}
	\tablehead{
		\colhead{Model} &
		\colhead{Type} &
		\colhead{Engine Type} &
		\colhead{\Ni\ Mass} &
		\colhead{Remnant} & 
		\colhead{$M_{ej}$}  
		\\
		\colhead{} &
		\colhead{} &
		\colhead{}    &
		\colhead{$\Ms$}    &
		\colhead{}   &
		\colhead{$\Ms$}  
	}
	
	\startdata
	A &  J & J($2.5^{\circ}$, $\epsilon$)  &  $0.016$ &   $4.54\Ms$ BH  & 5.46      \\
	B  & W &  W($\epsilon$) &  0.038&  $1.66 \Ms$ NS     & 8.34       \\
	C  &  J+W  &  J($2.5^{\circ}$, 0.1$\epsilon$), W(0.9$\epsilon$)    & 0.036 &  $1.66 \Ms$ NS &   8.34           \\
	D  &  J+W  &  J($10^{\circ}$, 0.1$\epsilon$), W(0.9$\epsilon$)  & 0.037   &  $ 1.66 \Ms$ NS    &   8.34   
	\enddata
	\tablecomments{The rate of energy deposition is $\epsilon = 10^{51}$ \erg\ s$^{-1}$ and its 
		duration is 20 sec. \ken{"J" stands for jet and "W" for wind, and this angle within "J" is its half-opening angle.}}
	\label{tbl:model}
\end{deluxetable*}


\begin{deluxetable*}{ccccc}
	\tablecaption{\Ni\ masses}
	\tablehead{
		\colhead{Energy Injection Zone} &
		\colhead{Mode A} &
		\colhead{Mode B} &
		\colhead{Mode C} &
		\colhead{Mode D} 
		\\
		\colhead{} &
		\colhead{\Ms} &
		\colhead{\Ms}    &
		\colhead{\Ms}    &
		\colhead{\Ms}   
	}
	
	\startdata
	2000 - 3000 km  &   0.024   &   0.036  &  0.037   &   0.036           \\
	2000 - 4000 km  &   0.015   &   0.040  &  0.039   &   0.040             \\
	2000 - 5000 km  &   0.016   &   0.038  &  0.036   &   0.037           
	\enddata
	\label{tbl:ni}
\end{deluxetable*}

\subsection{\STELLA}

We calculate light curves for our explosions with the 1D multigroup radiation hydrodynamics 
code \STELLA\ \citep{Bli93,Bli06,Bli11}. \STELLA\ can calculate spectral energy distributions 
(SEDs) for the blast profiles at each time step. Multicolor LCs can be obtained by convolving 
filter functions with the SEDs.  All our SN light curves are calculated with 100 frequency bins 
from 1 to $5\times10^4$  \AA\ on a log scale.  \STELLA\ implicitly evolves time-dependent 
equations of the angular moments of the intensity averaged over a frequency bin.  Local 
thermodynamic equilibrium is assumed in determining the ionization states of materials.  
\STELLA\ has been extensively used for modeling SN light curves \citep{Bli00,Chu04,Woo07,
Tom11,Mor11}. 

\section{Magnetar Powered Explosions}

The central engine injects $10^{51}$ erg/s into a shell of $r = 2000 - 5000$ km for 20 sec, 
heating the gas and producing a strong shock that reverses the collapse.  The energy is 
evenly distributed through the shell, and in models B, C, and D the heated gas reverses 
infall in the shell in under a second. Most of the \Ni\ forms in the first 0.5 sec of collapse, 
when the gas is heated to $4.3 \times 10^9$ K and compressed to $\rho > 10^6$ \gcc\ by 
energy from the magnetar below and by infall from above.  In models B - D, most of the 
\Ni\ is formed by compression heating and later expelled by the magnetar wind.  0.026 - 
0.04 \Ms\ of \Ni\ are ejected by the explosion, which drives a strong shock from the core 
at 1 - 4 $\times 10^9$ cm s$^{-1}$.  Energy from the magnetar primarily governs the 
dynamics of the explosion, not energy from nuclear burning.  

\Ni\ masses are listed in Table~\ref{tbl:model} for all four models.  The isotropic models 
create more \Ni\ than the jet model, in which material is burned in a much smaller solid 
angle.  Nevertheless, the jet produces more \Ni\ than its small solid angle alone might 
suggest because it dredges material up from greater depths that would otherwise have 
fallen onto the compact remnant.  Models B, C and D produce about the same \Ni\ 
mass, 0.037 \Ms, which is significantly less than would be expected for a CC SN for 
reasons we discuss in greater detail below.

\subsection{\Ni\ Production}

Distributing the energy of the magnetar evenly from radii of 2000 - 5000 km is reasonable 
but {\em ad hoc}.  To determine how sensitive \Ni\ production is to the thickness of the 
injection site we repeated models A - D at higher resolution on much smaller meshes. The 
grid was reduced to $10^{10}$ cm along both axes, with $512^2$ zones and no AMR.  We
considered three injection regions:  2000 - 3000 km, 2000 - 4000 km and 2000 - 5000 km.  
Although energy from the magnetar is deposited for 20 sec, densities at the injection site
fall very quickly so no \Ni\ will be made after 0.5 sec after the explosion. We therefore only 
evolve the SN for the first 5 sec in these models.  The outcomes of all twelve runs are 
summarized in Table \ref{tbl:ni}. 

\Ni\ masses are sensitive to the structure of the progenitor for two reasons.  First, the 
structure determines how much mass is burned to \Ni\ by the magnetar wind and by 
compression heating from collapse.  It also determines how much of this \Ni\ then falls 
back onto the compact remnant.  We show the structure of the presupernova core in 
\Fig{p_ni}.  Gas below $r \sim 1.57 \times 10^8$ cm (the radius of the iron core) will 
collapse directly to a proto-neutron star.  Most of the \Ni\ forms at radii of $\sim 1.57 - 
2.38 \times 10^8$ cm, where densities and temperatures are driven to $5.7 - 25 \times 
10^6$ \gcc\ and $3.07 - 4.64  \times 10^9$ K by the wind and by collapse.  If the 
magnetar forms promptly, the gas in this shell would be burned to 0.21 \Ms\ of \Ni, 
about what would be expected from a CC SN.  

But in most scenarios the magnetar does not turn on immediately after proto-neutron star 
formation.  If the delay is just 0.2 sec, \Ni\ below $r \sim 1.85 \times 10^8$ cm would fall 
back onto the NS and only 0.05 - 0.12 \Ms\ would be ejected.  If the delay is 0.5 sec, \Ni\ 
at $r \le 2.0 \times 10^8$ cm will fall back and even less \Ni\ will be ejected, 0.01 - 0.07 
\Ms.  Our choice of energy deposition at $r \ge 2 \times 10^8$ cm implies a 0.5 sec delay 
in the explosion, which is why on average only 0.037 \Ms\ of \Ni\ is produced in models B
- D.

The reason why \Ni\ production is not sensitive to the thickness of the injection region is 
that no \Ni\ is formed at $r >$ 2300 km in our models.  We again find that the jet in model
A dredges up more \Ni\ from lower radii than it forms on its own. The energy was injected 
in a half-opening angle of only $2.5^\circ$. If all the mass within this solid angle was 
burned to \Ni\ in the $r = 2000 - 3000 / 4000$ km tests, it would only make $1.65 \times 
10^{-3}$ \Ms, not the 0.015 - 0.016 \Ms\ it actually dredges up.  

The extra material dredged up by the jet in model A is due to lateral pressure forces that
drive gas sideways out of the jet when energy is suddenly injected into its solid angle.
The gas that is driven sideways also experiences shear forces upward from gas deeper 
within the solid angle that is mostly expanding outward along the axis of the jet, and the 
combination of the two motions is what brings up the extra material. In the Model A runs, 
the simulation with magnetar energy injection at $r = 2000 - 3000$ km produces 0.024 
\Ms\ of \Ni, nearly twice that of the other two sites, because the lateral pressure forces 
are much stronger when the injected energy is concentrated in the thinner shell, and 
they drive more vigorous sideways mixing.

Our choice of injection site is consistent with simple analytic estimates of the radius out 
to which material can be burned to \Ni\ in the core.  The hydrodynamic time of the core 
is
\begin{equation}
t_{dyn} = \frac{446}{\sqrt{\rho_{\rm{c}}}} \sec,
\end{equation}
where $\rho_{\rm{c}}$ is its mean density. If we take $\rho_{\rm{c}} = 10^6$ \gcc, then 
$t_{dyn} \sim 0.5$ sec, which is when most explosive burning and \Ni\ production would 
happen.  Over this time the magnetar injects $0.5 \times 10^{51}$ \erg\ into the outer 
layers of the core, and most of this energy goes into explosive burning.  Except at small 
radii near the origin of the shock, the peak temperature at radius $r$ can be obtained by 
setting $(4/3) r^3aT_{\rm{s}}^4 \sim E_{\rm{exp}}$, where $T_{\rm{s}}$ is the shock 
temperature and $a$ is the radiation density constant.  This equation assumes that the 
heat capacity of the material behind the shock is dominated by the radiation field, and 
that expansion and pressure waves behind the shock are capable of maintaining nearly 
isothermal conditions there. The shock temperature at radius $r$ can then be expressed 
as \citep{Woo02} 
\begin{equation}
T_{\rm{s}}(r) = 1.33\times 10^{10} \left( \frac{E_{\rm{exp}}}{10^{51} \erg} \right)^{1/4} \left( 
\frac{r}{10^8 \cm}  \right)^{-3/4}  {\rm K},
\end{equation}
In our simulation, $E_{\rm{exp}}=0.5 \times 10^{51}$, so a spherical volume with $r = 2.91 
\times 10^8 \cm$ will reach $T_s \sim 4.5 \times 10^9$ K for \Ni\ synthesis, which is 
consistent with our \CASTRO\ models.


\begin{figure}[h]
	\begin{center}
		\includegraphics[width=\columnwidth]{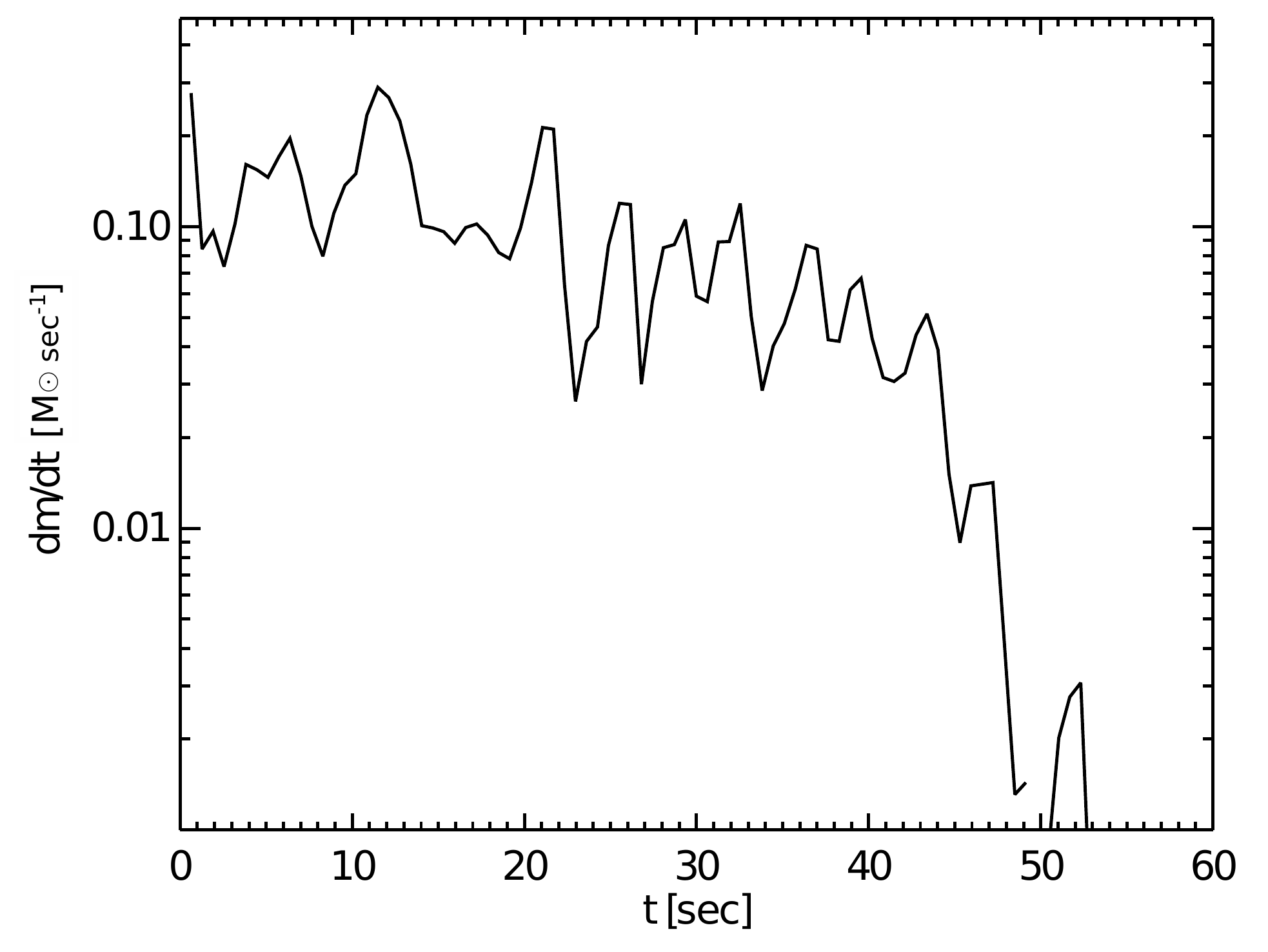} 
		\caption{Accretion history in model A.  Accretion proceeds for 60 sec after the launch of 
			the jet.  The rates vary from 0.01 -- 0.3 \Ms\ s$^{-1}$. Sound waves created by accretion 
			flows have a velocity of $\sim 10^8 \cm \sec^{-1}$ and cause the oscillations.  Accretion 
			eventually turns the NS into a black hole. \label{fig:acc}}
	\end{center}
\end{figure}


\begin{figure}[h]
	\begin{center}
		\includegraphics[width=0.8\columnwidth]{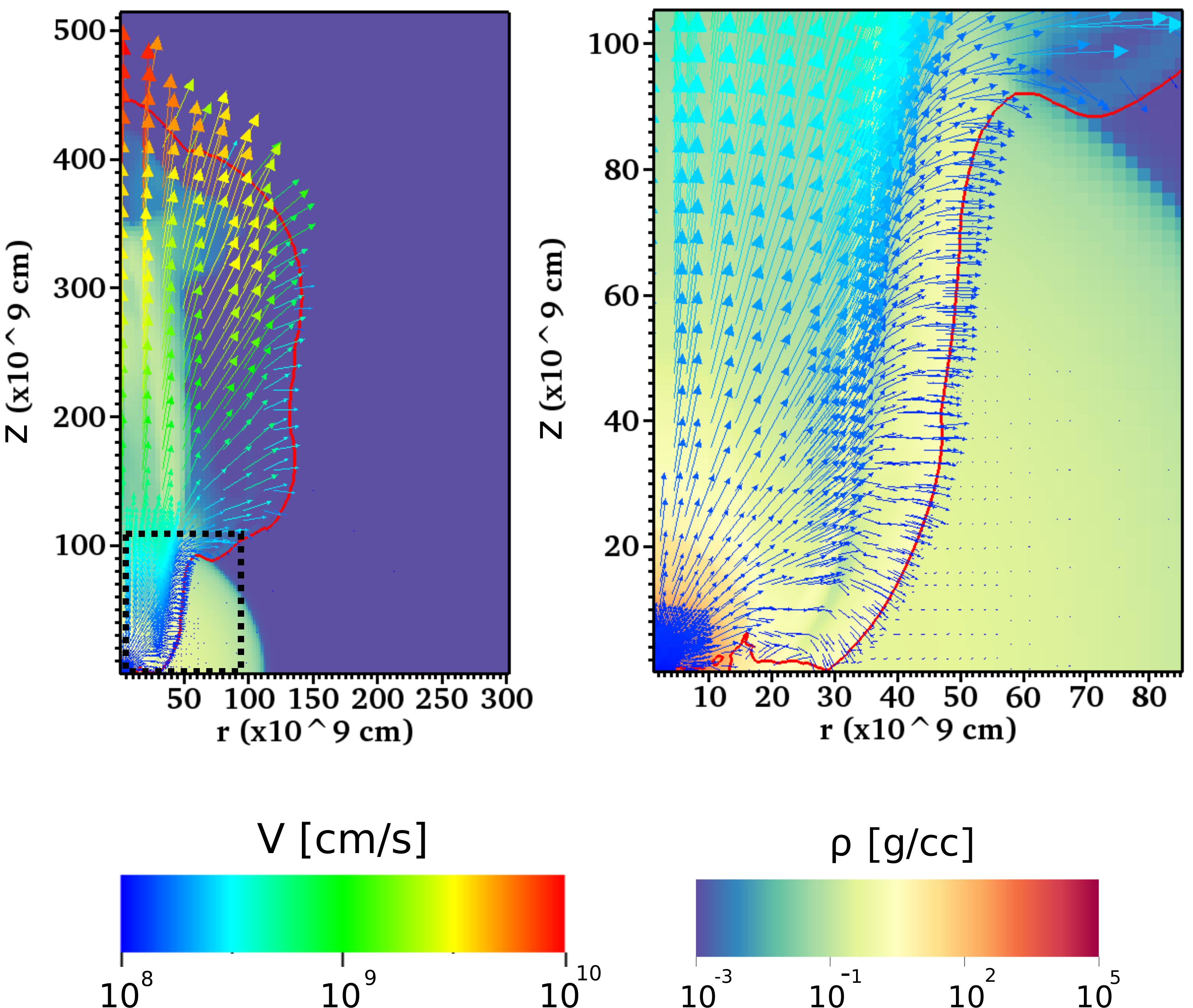} 
		\caption{\ken{Outflow driven by the jet in model A at about 100 sec (left panel). Right panel shows the close-up of the dashed-line box in left panel. 
				Vectors show the velocities of gas and color map shows the densities. The red line marks the boundary of jet-heated gas.
				There is a distinct broadening effect of jet at the stellar surface.  Because many fine AMR zones are  generated for a high density region, it causes concentration of velocity arrows around the center of star.  Gas heated by the jet blows 
				out a fraction of the star and eventually halts accretion flow from the equator.  There are fluid instabilities developing around of the equatorial plan between $r \approx 10^{10} - 3 \times 10^{10}$ cm where cocoon hit the boundary of simulation domain. In a full-star simulation, more violate fluid instabilities may emerge when cocoons from the upper and  lower jet collide each other at the equatorial plane.  }   
			\label{fig:infall}}
	\end{center}
\end{figure}

\subsection{Accretion and Breakout}

The jet in model A gradually broadens as it propagates through the star at $\sim$ 15\% of 
the speed of light, breaking through its surface at 19 sec.  The shock driven by the wind 
in model D breaks out at 73 sec and the shock driven by both a wind and jet breaks out 
of the star at 38 and 57 sec in models C and D.  In model A, a small fraction of the 0.016 
\Ms\ of the \Ni\ ejected by the jet is present in the bow shock.  Since the jet propagates 
mostly along the axis of the star, accretion continues inward along the equatorial plane. 
We show accretion rates and blowout for model A in \Fig{acc} and \Fig{infall}.  \ken{
We calculate the mass accretion by assuming the mass flux across a surface at radius 
of  $r = 2\times10^8$ cm. However, this accretion is subject to fluid instabilities. At $t= 50$ sec, 
two data points are missing because accretion flux becomes an outflow due to the strong fluid instabilities  before the accretion shuts down.} Accretion 
persists for 52 sec in model A until it is completely shut down by the cocoon of hot gas 
from the jet.  The rate varies from 0.01 - 0.3 \Ms\ s$^{-1}$ and the total accreted mass is 
3.03 \Ms, enough to collapse the NS to a 4.54 \Ms\ black hole (BH). From \Fig{acc}, we estimate 
the BH forms about 10 sec after accretion starts when the mass of central compact object exceeds 2 \Ms.  Energy injection by the jet may become less efficient after 
the BH forms, but exactly how the NS collapses to a BH is beyond the scope of our study.  
While we just assume that the mass that falls back onto the NS turns it into a BH, angular 
momentum transport can delay BH formation and radiation from the BH may then reduce 
accretion.  In \Fig{infall} the core of the star has been completely blown out along the axis 
of the jet but remains intact along the equatorial plane, and likely falls back onto the BH at 
later times.

\subsection{Mixing}

\begin{figure}[h]
 	\begin{center}
 		\includegraphics[width=.8\columnwidth]{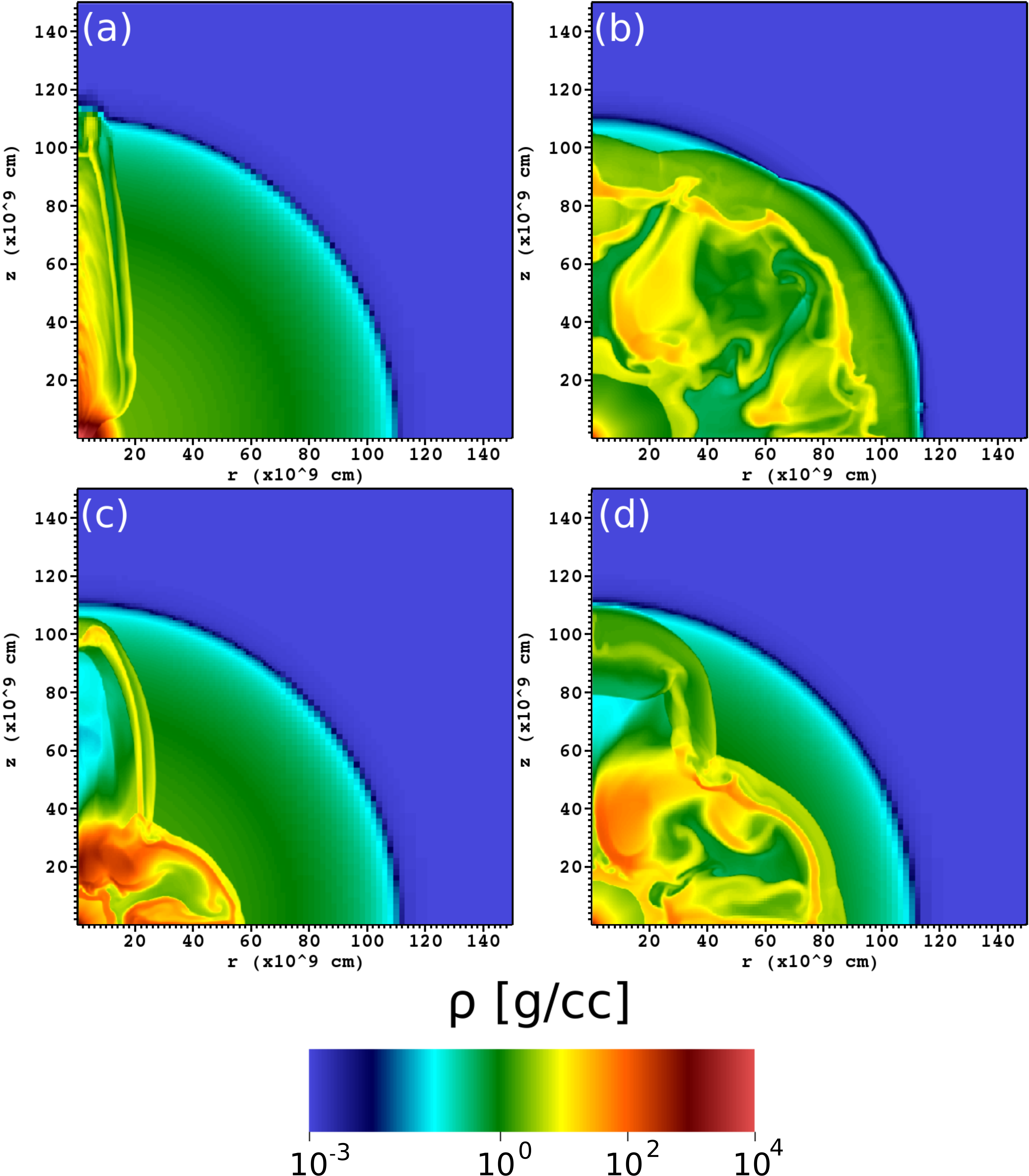} 
 		\caption{Panels (a) - (d) are density images of models A - D at shock breakout \ken{at time = 19, 73, 38, and 57  sec.}  Different explosion engines already produce distinctive mixing at early times.  \label{fig:breaks}}
 	\end{center}
 \end{figure}

Although all four models have the same injected energy, the geometry of injection leads to 
a range of mixing during the explosion.  We show the four SNe at shock breakout in 
\Fig{breaks}.  In model A the jet produces a strong collimated outflow in the polar direction. 
The flow exhibits knots and kinks due to jet instabilities and some shear instabilities, but the 
latter do not grow to sufficient amplitudes to produce much mixing before breakout.  Some 
heavy elements deep in the star are dredged up by the jet.  Model B exhibits much more 
mixing, on par with what would be expected for a CC SN.  \ken{The collision between the wind 
and the collapsing outer layers of the star drives the dynamical instabilities. The formation 
of a reverse shock also drives Rayleigh-Taylor (RT) insatiabilities and contributes to mixing significantly}.   
In models C 
and D, twisting in the jet and the disk wind leads to both shear and RT 
instabilities, as seen in the elongated structures in the ejecta.  There are traces of \Ni\ in 
the jet, so gamma-ray emission from \Ni\ is possible at early phases of the explosion. In all 
four cases the star becomes unbound by 80 sec after the explosion, but mixing continues 
after breakout. 

The CSM around massive stars is usually diffuse because their strong winds and large 
ionizing UV fluxes drive away gas in their vicinity.  The shock therefore accelerates rapidly 
when it breaks out of the star.  The dramatic drop in density between the surface of the 
star and the CSM sometimes crashed the hydro solver in our simulations \citep[a problem 
that has been reported by others; see, e.g.,][]{Wha12c,Wha12b}.  To prevent numerical
difficulties we take the CSM to fall off from the surface of the star as $\rho \propto r^{-3.1}$.  
This diffuse envelope also prevents the formation of reverse shocks in the ejecta that can 
cause additional mixing after breakout so we can study just the mixing that is intrinsic to 
the explosion itself.  In model A the velocity of the jet at breakout is $\sim$ 7 $\times$ 10$
^9$ cm s$^{-1}$, or $\sim$ 25\% of the speed of light, so it is mildly relativistic.  The 
breakout velocity is sensitive to the CSM density so if it falls off dramatically the shock can 
accelerate to nearly the speed of light and produce a GRB, although this does not happen
in our models.  

\begin{figure}[h]
 	\begin{center}
 		\includegraphics[width=.8\columnwidth]{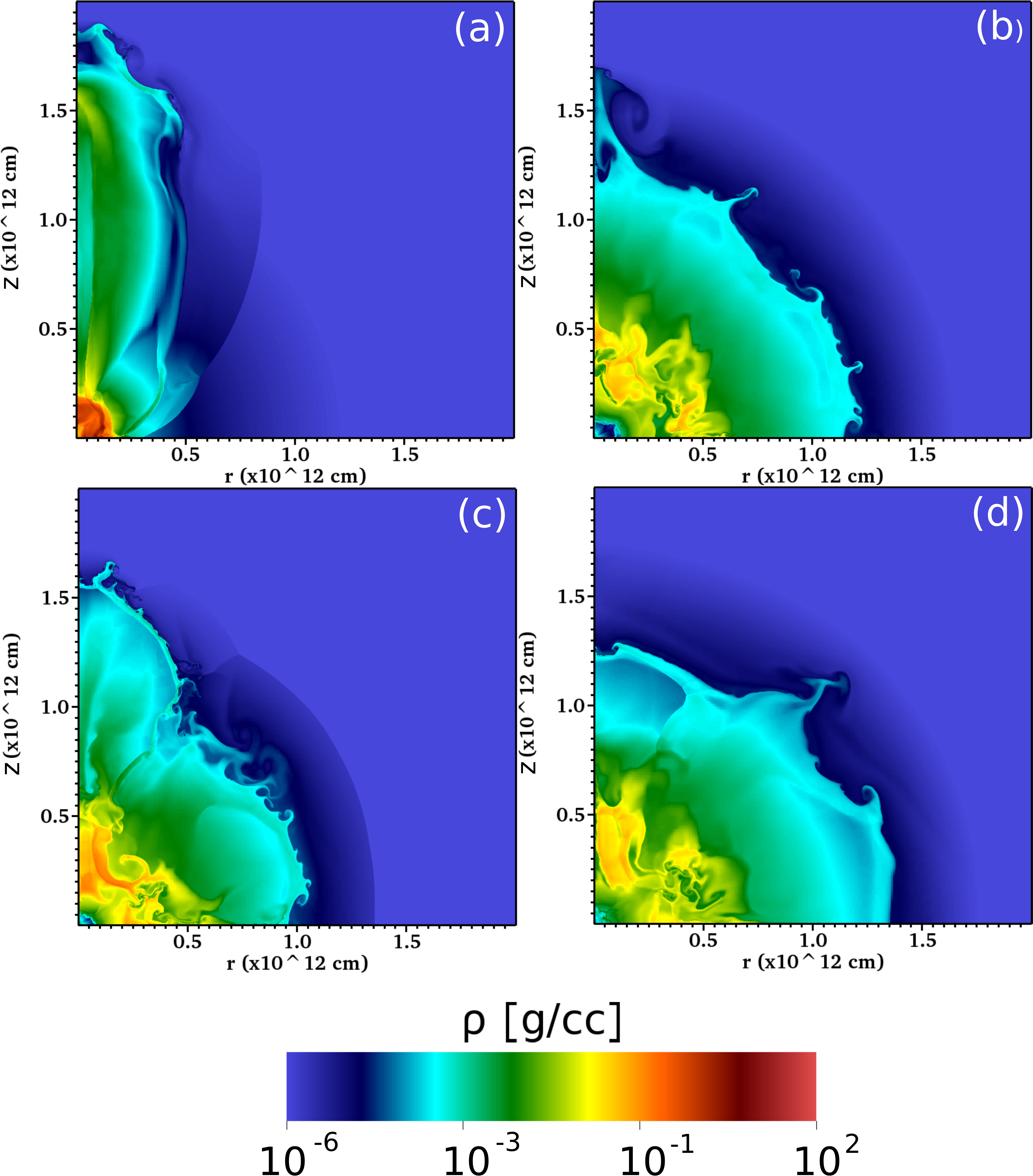} 
 		\caption{Panels (a) - (d) are density images of models A - D at the end of the simulations, 
 			when the shock has reached a radius of about 15 times the radius of the star \ken{at time = 293, 414, 354, and 458 sec}. At this stage, the explosion has successfully unbind the entire 
 			star and mixing is getting frozen.   
 			\label{fig:breakd}}
 	\end{center}
 \end{figure}

We run our simulations until the shock or jet reaches $\sim$ 20 $r_*$ and mixing ceases.  
At this stage the ejecta are expanding homologously and their energy is almost entirely 
kinetic (their internal energy is $\sim$ 10\% of the total energy).  We show snapshots of 
the ejecta for all four models at this stage in \Fig{breakd}.  The jet continues to broaden 
well after breakout. The original opening angle of $5^\circ$ has grown to $46^\circ$ as it 
partially thermalizes with gas in its path, forming a shock that is perpendicular to the jet.  
Models B and D have similar morphologies. Jets with large opening angles soon become 
hard to distinguish from explosions driven just by disk winds because the energy of the 
jet is rapidly spread throughout the ejecta. 
 
\begin{figure*}
 	\begin{center}
 		\includegraphics[width=\textwidth]{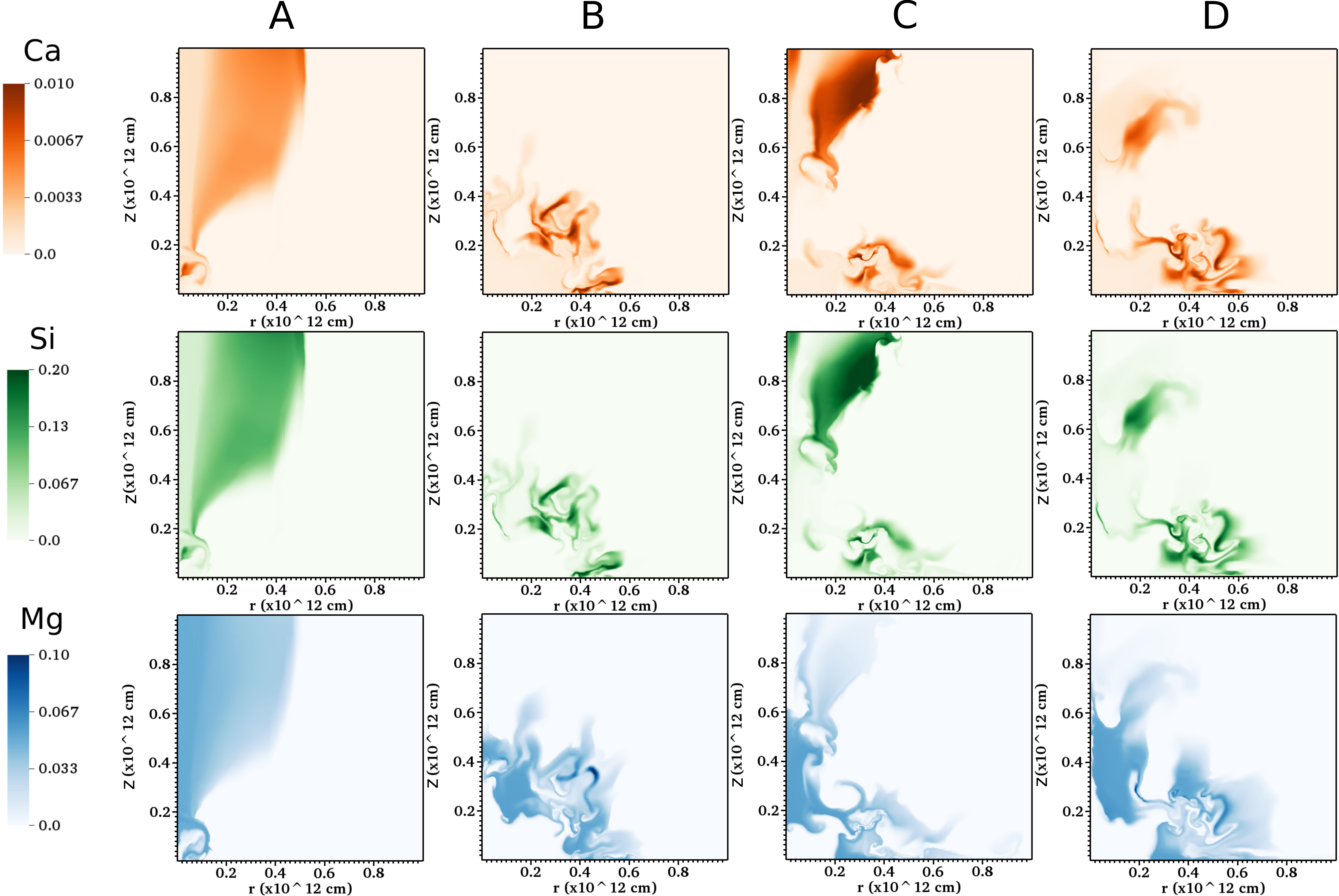} 
 		\caption{Images of \Mg, \Si, and \Ca\ in the ejecta of the four explosions at the end of the 
 			simulation.  These species are in the unit of mass fraction. In model A the metals mostly trace the outline of the jet but in model 
 			B their distribution is more like that in normal CC SNe. The distribution of metals in models 
 			C and D are intermediate to those in A and B. 
 			\label{fig:iso}}
 		\vspace{0.3in}
 	\end{center}
 \end{figure*}

We summarize mixing in all four models in \Fig{iso}. The metals mostly trace the outline 
of the jet in model A, although there is clearly some mixing.  A ring pattern created by the 
cocoon of the jet is also visible.  There is very little mixing along the equatorial plane of 
the star.  There is much more mixing in models B, C, and D.  The shells of elements that 
build up inside the star over its life are completely disrupted by the shock.  Models B and 
D, which are somewhat difficult to distinguish in density, are more easily differentiated by 
isotope distribution, which can provide a diagnostic of central engine type which can be 
examined in SN spectra.
 
 
\section{Light Curves}
 
To calculate light curves for these explosions we average the flow variables at each radius 
over a sample of 10 angles on the grid.  We then map these angle-averaged radial profiles 
onto 1D spherical grids in the \STELLA\ code.  The blast profiles are taken at the onset of 
homologous expansion. We only calculate light curves for models B, C, and D because the 
jet in model A is highly directional and relativistic so a different technique would be required 
to compute its light curve. We evolve all three explosions out to 60 days.

\begin{figure}
	\begin{center}
		\includegraphics[width=.8\columnwidth]{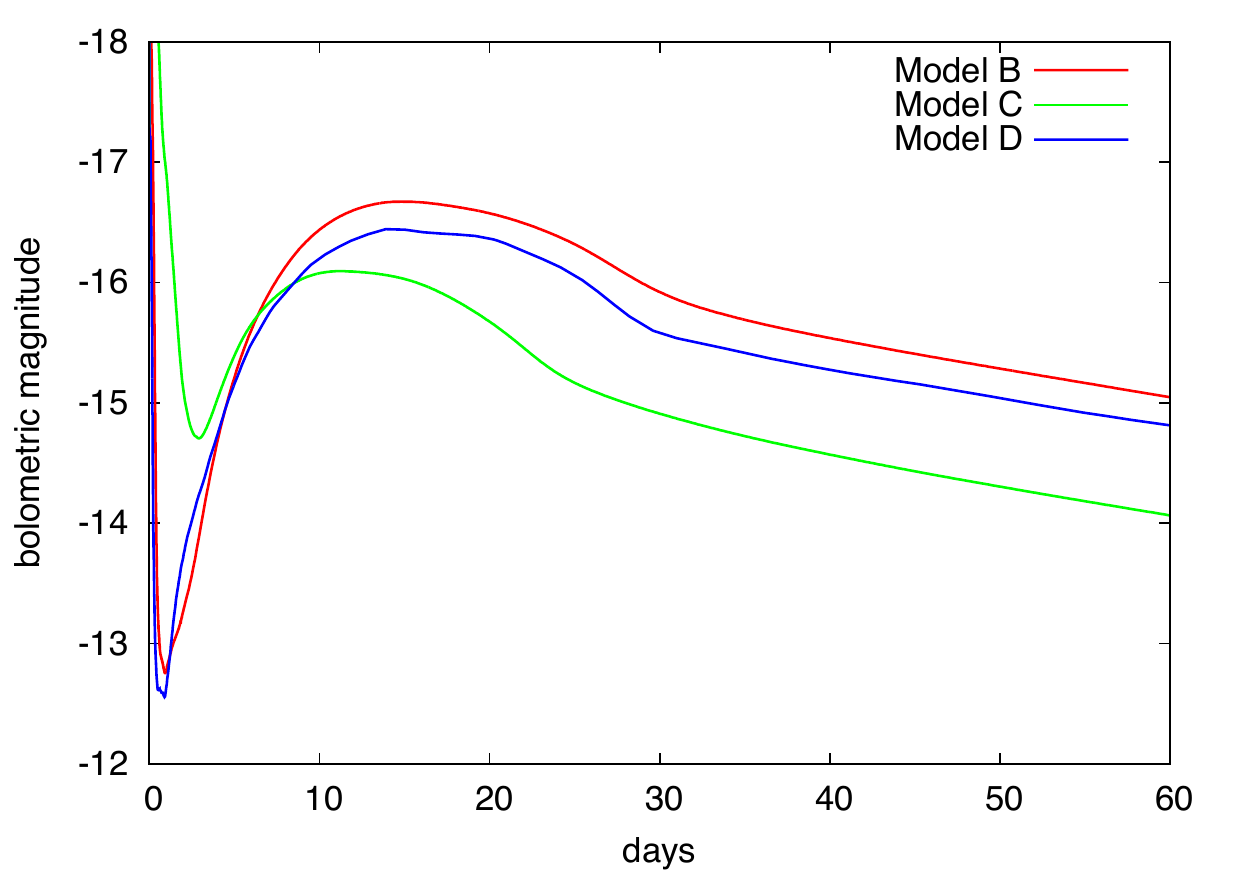} 
		\caption{Bolometric light curves for models B, C, and D. A very luminous breakout transient 
			that lasts for about an hour appears first and then dims as the fireball expands and cools.  
			The light curves then rebrighten as photons from \Ni\ decay begin to diffuse out of the ejecta, 
			reaching peak $M_b \approx -16.5$ about 15 - 20 days after the explosion. \Co\ decay follows, 
			and the the light curves gradually fade on timescales comparable to its half-life of 77 days. 
			\label{fig:lc1}}
	\end{center}
\end{figure}

Bolometric light curves for the three models are shown in \Fig{lc1}.  Each exhibits a brief, 
extremely luminous pulse due to shock breakout that lasts for about an hour. The flux then 
dims and later rebrightens as photons from the radioactive decay of \Ni\ begin to diffuse 
out of the ejecta, which occurs on a timescale $t_{\mathrm{diff}}$ \citep{Arn96}, where
\begin{equation}
t_{\mathrm{diff}} \sim 14 \hspace{0.05in }\kappa^{\frac{1}{2}} M_{\mathrm{ej}}^{\frac{3}{4}} 
E^{-\frac{1}{4}}   \hspace{0.2in }  \mathrm{days} 
\end{equation}
and $\kappa$ is the opacity of the ejecta in units of the Thompson scattering opacity of 
free electrons in hydrogen-free  gas (0.2 cm$^2$ g$^{-1}$), $M_{\mathrm{ej}}$ is the 
mass of the ejecta through which photons must diffuse to reach the surface in units of 
solar mass and $E$ is the energy of the explosion in units of  10$^{51}$ erg.  Taking 
$\kappa =$ 1, $M_{\mathrm{ej}} =$ 4 because of the partial dredging up of \Ni, and $E 
=$ 20, we find that $t_{\mathrm{diff}} \sim$ 18 days, which corresponds to the time at 
which \Ni\ rebrightening peaks in all three magnetar-powered SNe. \Co\ decay follows, 
and the light curves gradually fade on timescales comparable to its half-life, $\sim$ 77 
days.

\begin{figure}
	\begin{center}
		\includegraphics[width=\columnwidth]{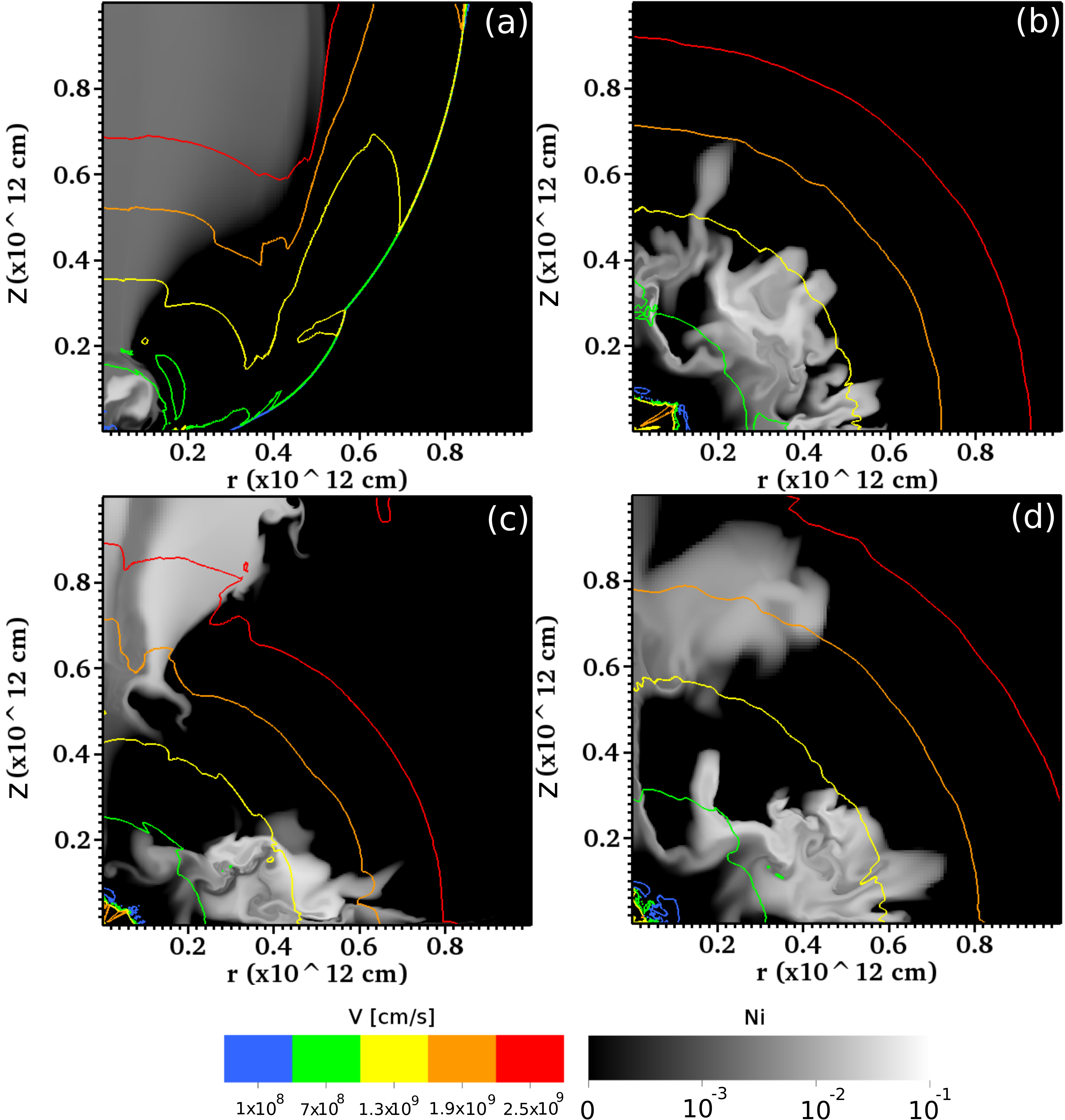} 
		\caption{\Ni\ mass fractions and radial ejecta velocities. Clumps of \Ni\ have broken 
			through the ejecta along the axis of the jet in models A, C and D, but not in B, in which there 
			is no jet. The roughly evenly spaced velocity contours suggest homologous expansion.
			\label{fig:nivel2}}
	\end{center}
\end{figure}

Because less than 0.05 \Ms\ of \Ni\ is made, bolometric luminosities peak at $M_b \approx 
-16.5$, which is not much brighter than normal CC SNe. Mixing may dredge \Ni\ up to nearly 
the surface of the ejecta in some models.  We show  \Ni\ mass fractions and velocities at 
the end of all four runs in \Fig{nivel2}.  Except in model B, which is driven by only a wind, 
a clump of \Ni\ has broken through the ejecta along the axis of the jet.  The \Ni\ rich ejecta 
have velocities of $1 - 3 \times 10^9$ cm s$^{-1}$ that are large enough to Doppler shift 
their spectra.  Note the roughly evenly spaced velocity contours, which indicate 
homologous expansion of the ejecta.

Model A could produce a GRB without a luminous SN component.  If the jet is sustained by 
the accretion of gas falling in from the equator, it could drive a GRB from 20 s to 100 sec. 
This model could explain GRB 060614 \citep{Gal06, Del06}, which synthesized very little 
\Ni, $5 \times 10^{-4}$ \Ms, lasted $\sim 100$ s, and did not produce an optically luminous 
SN ($M_v > -12.3$ mag).  


\section{Summary and Conclusions}

We have performed 2D simulations of magnetar-powered SNe with the \CASTRO{} AMR 
code.  These transients are presumed to be observed as SNe Ic-BL.  We examine central
engines in the form of disk winds, jets, and combinations thereof by varying the morphology 
of energy injection in the core of the star.  These engines lead to a variety of mixing in the 
explosion before and after shock breakout.  We find that infall from the equatorial plane can 
create a BH if there is a collimated jet; otherwise, a NS forms.  

Although magnetar-powered SNe can be ten times more energetic than normal CC SNe, as 
is observed with SNe~Ic-BL, they do not necessarily make superluminous transients.  Light 
curves for our SNe Ic-BL manifest a sharp, very luminous transient due to shock breakout 
followed by dimming and then mild \Ni\ rebrightening at 10 - 20 days. Our SNe Ic-BL are not 
very luminous because they only make 0.02 - 0.05 \Ms\ of \Ni, which is consistent with 
models by \citet{Nis15} and \citet{Suw15}.  The \Ni\ production is small because most of it 
is made by compression heating during the core collapse instead of energy injection from 
the magnetar, and much of it falls onto the nascent NS before it can be expelled by the 
magnetar wind. Unless energy is deposited at a very high rate, the magnetar is unlikely to 
produce enough \Ni\ to rebrighten its light curve. \citet{Suw15} suggests that magnetars 
with magnetic fields $> 2  \times 10^{16}$ G and rotation periods $<$ 1 ms (energy 
deposition rates $> 4 \times 10^{51}$ erg s$^{-1}$) can produce  $>$ 0.2 \Ms\ \Ni.  

However, extreme magnetars such as these may release a significant fraction of their 
rotational energy as gravitational waves rather than dipole radiation, so their \Ni\ yields may 
still be low.  But these events could become much brighter if their ejecta crash into a dense 
CSM \citep[e.g.,][]{Mor13,Wha12e}.  Furthermore, if a successful CC explosion is followed 
by the formation of a magnetar with magnetic fields of 10$^{14}$ - 10$^{15}$ G and rotation 
periods $\sim$ 5 ms, it may become very bright at later times \citep{Woo10,Kas10,Che16}. 
\Ni\ blown out by jets at high velocities might also emit observable gamma rays.

In future models the nuclear reaction network could be improved to calculate better yields 
for magnetar-powered SNe.  More realistic initial conditions for the core collapse engine 
based on dedicated 2D or 3D simulations rather than energy injection by hand could also 
be implemented.  Although we do not include radiation transport our results indicate that
multidimensional radiation hydrodynamics will be required to determine how photons are 
emitted from dense structures created by fluid instabilities \citep{Che14, Che16} and 
produce more accurate light curves.  More realistic prescriptions for the CSM can also be 
calculated from radiation hydrodynamical models of the ambient H II region and wind cavity 
of the star.  In such profiles, the jet might reach velocities of 0.95$c$ and produce a GRB, 
mandating special relativistic upgrades to the hydro solver in \CASTRO.  Another scenario 
worth consideration is the collision of the asymmetric ejecta with a CSM. 

Our models will soon be confronted by more detections of magnetar-powered SNe in SN 
searches with PTF, Pan-STARRS, LSST and future searches with {\em Euclid} and the 
Wide-Field Infrared Survey Telescope (WFIRST).  These exotic explosions may have 
occurred more frequently in the primeval universe because the Population III initial mass 
function (IMF) may have been top-heavy \citep[e.g.,][]{Fsg09,Wha12,Glo12}.  Many 
primordial stars may also have been born with rotation speeds close to the breakup limit 
\citep{Sta11b, Sta13} or in binaries \citep{Tur09,Sta10,Sb13}.  Some of these highly 
energetic transients \citep[e..g.][]{Nak12} may also be found in the near infrared by the {\it 
James Webb Space Telescope} and ground-based 30m telescopes and probe the 
properties of the first stars in the Universe.

\acknowledgements 

\ken{We thank the anonymous referee, whose comments and suggestions improved the quality of this paper}. The authors also thank Ann Almgren, Weiqun Zhang, and Sergei Blinnikov for technical support 
with \CASTRO\ and \STELLA.  K.C. acknowledges the support of an EACOA Fellowship 
from the East Asian Core Observatories Association and the hospitality of the Aspen Center 
for Physics, which is supported by National Science Foundation grant PHY-1066293.  Work 
at UCSC was supported by an IAU-Gruber Fellowship, the DOE HEP Program 
(DE-SC0010676) and the NASA Theory Program (NNX14AH34G).  D.J.W. was supported 
by the European Research Council under the European Community's Seventh Framework 
Programme (FP7/2007-2013) via the ERC Advanced Grant "STARLIGHT: Formation of the 
First Stars" (project number 339177). YS was supported in part by the Grant-in-Aid for 
Scientific Research (Nos. 16K17665 and 16H00869). V.B. acknowledges support from NSF grant AST-1413501.  \CASTRO{} was 
developed through the DOE SciDAC program by grants DE-AC02-05CH11231 and DE-
FC02-09ER41618.  Our numerical simulations were performed at NERSC and the Center 
for Computational Astrophysics (CfCA) at the National Astronomical Observatory of Japan 
(NAOJ).

\bibliographystyle{apj2}

\end{document}